\documentclass[final]{main}

\usepackage{times}
\usepackage{epsfig}
\usepackage{graphicx}
\usepackage{amsmath}
\usepackage{amssymb}
\usepackage{dsfont}
\usepackage{gensymb}
\usepackage{multirow}

\usepackage[pagebackref=true,breaklinks=true,colorlinks,bookmarks=false]{hyperref}

\def\cvprPaperID{7786} 
\def\confYear{CVPR 2021}

\begin{document}

\title{3D Scattering Tomography by Deep Learning\\with Architecture Tailored to Cloud Fields}
\author{Yael Sde-Chen, Yoav Y.~Schechner, Vadim Holodovsky\\
Viterbi Faculty of Electrical Engineering\\ 
Technion - Israel Institute of Technology \\
Haifa, Israel\\
{\tt\small yael.sde.chen@gmail.com}, {\tt\small yoav@ee.technion.ac.il}, {\tt\small vholod@ef.technion.ac.il}
\and
Eshkol Eytan\\
Department of Earth and Planetary Science \\
The Weizmann Institute of Science\\ 
Rehovot, Israel \\
{\tt\small eshkol.eytan@weizmann.ac.il}
}

\maketitle
\begin{abstract}
We present 3DeepCT, a deep neural network for computed tomography, which performs 3D reconstruction of scattering volumes from multi-view images. 
Our architecture is dictated by the stationary nature of atmospheric cloud fields.
The task of volumetric scattering tomography aims at recovering a volume from its 2D projections. This problem has been studied extensively, leading, to diverse inverse methods based on signal processing and physics models. However, such techniques are typically iterative, exhibiting high computational load and long convergence time.
We show that 3DeepCT outperforms physics-based inverse scattering methods in term of accuracy as well as offering a significant orders of magnitude improvement in computational time. To further improve the recovery accuracy, we introduce a hybrid model that combines 3DeepCT and physics-based method. The resultant hybrid technique enjoys fast inference time and improved recovery performance.

\end{abstract}
\section{Introduction}
Artificial neural networks offer significant advantages to computer vision. In low-level vision, they have mainly progressed in addressing two-dimensional (2D) spatial fields. There is increasing effort to make such networks reconstruct three dimensional (3D) shapes or projections of opaque objects, either using explicit outer-shell or volumetric representations~\cite{choy20163d,chen2019learning,girdhar2016learning,wu2016learning,yan2016perspective,cciccek20163d,cao2018learning,xu2019disn,grant2016deep,xie2019pix2vox,johnston2017scaling,mescheder2019occupancy,riegler2017octnet,dou20173d,lombardi2019neural,mildenhall2020nerf}. However, there is still a significant gap in advancing reconstruction of 3D {\em heterogeneous volumetric translucent} fields, such as the atmosphere.

\begin{figure}[t]
\begin{center}
   \includegraphics[trim=11cm 5.7cm 3cm 0.5cm,clip, width=1\linewidth]{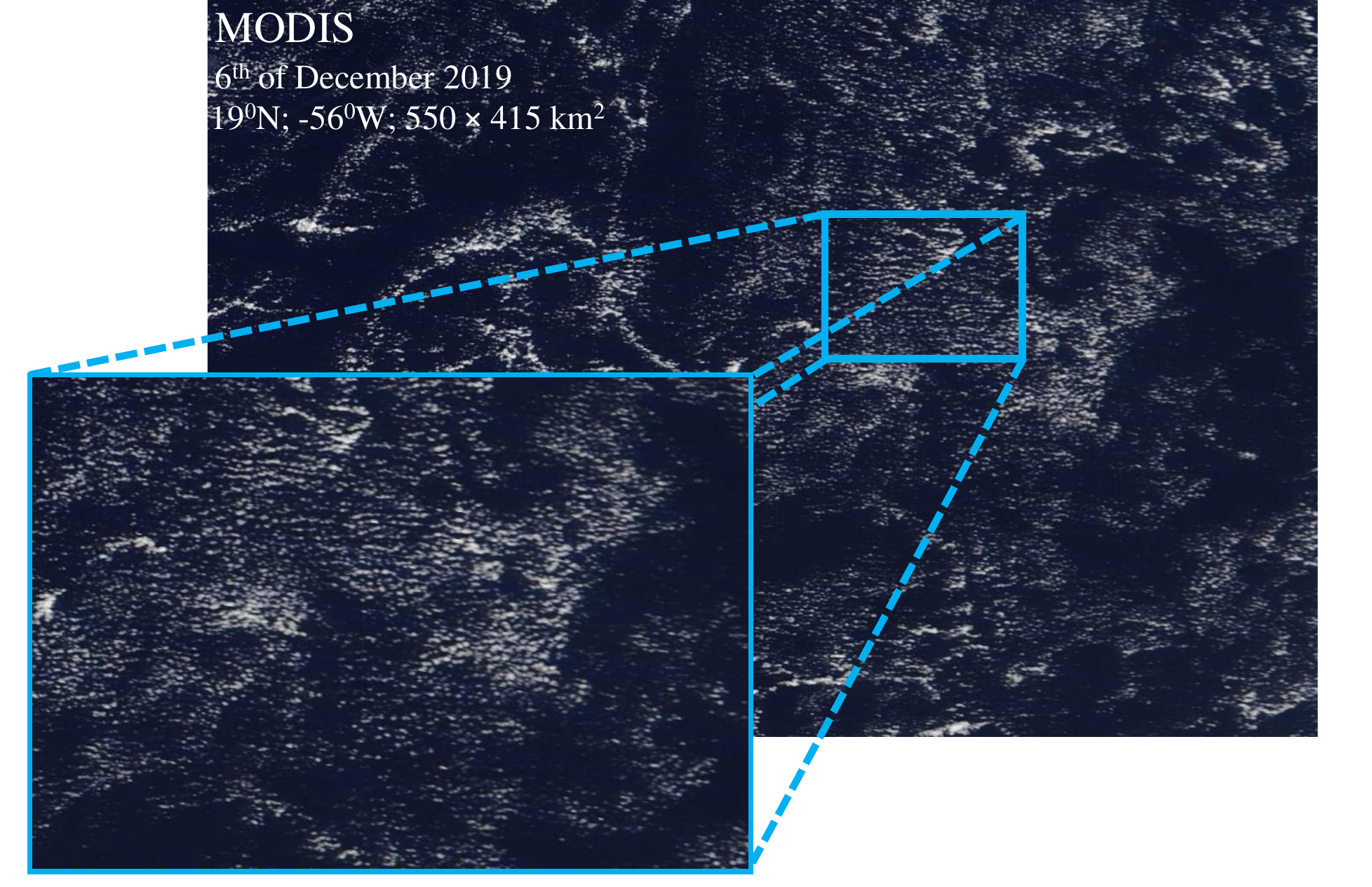}
\end{center}
   \caption{Shallow cumulus cloud fields. Terra satellite from \cite{worldview}.
}
\label{fig:example_cloud_field}
\end{figure}

3D heterogeneous translucent objects are reconstructed, essentially, using tomographic data. Radiation that propagates through the medium yields a set of multi-view 2D radiometric images. Analysis retrieves from these images a volumetric spatial distribution of material density in 3D. This is computed tomography (CT). It is used extensively in biomedical imaging and earth sciences~\cite{geva2018x,ren2020multiple,mejia2018cloud,boublil2015spatially,baruchel2000x,kazahaya2008computed}. Due to the critical role of CT in science, technology and medicine, it forms an important frontier in computer vision and computational photography research.

There are works that use deep neural networks (DNNs) to advance medical CT~\cite{wu2019computationally,qiu2019automatic,guo2020sensorless,shen2019patient,jnawali2018deep}. Nevertheless, most CT modalities are based on a linear image formation model, hence can be solved well using established signal processing methods, including optimization of a convex functional, without requiring a learning-based approach. Actually, it may be argued that 3D tomography does not lend itself easily to current DNNs. The reason is that DNNs require a lot of data to train, and it is extremely difficult to obtain sufficient ground-truth data of volumetric heterogeneous translucent objects. 

Reconstruction of such objects poses a serious challenge which is worth overcoming by DNNs. We believe this challenge and opportunity occur when these conditions are met. (1) The tomographic model is very complex: nonlinear, not unimodal. Then, classic methods of linear-CT analysis cannot apply. Moreover, optimization-based estimation is very slow, unscalable and too dependent on initialization.
(2) Scalability is critically needed to analyze huge 3D fields. (3) While the imaging model is nonlinear, it is continuous: an infinitesimal change of the medium continuously affects the image data, and vice versa. (4) There is a physics-based way to generate a large and diverse database.

If these conditions are met, then a DNN can realistically learn to express the richness of large translucent fields, and the physical processes that generate both 3D translucent objects and their images. Moreover, the inference speed offered by a trained DNN can significantly overtake explicit physics-based optimization. 

We pose a problem that should greatly benefit from a DNN for CT. The problem is imaging of a very large random 3D spatially heterogeneous {\em scattering} medium~\cite{martins2018harp,che2020towards,levis2017multiple,akkaynak2017space,gkioulekas2016evaluation,singh2017exploiting,satat2016all}: the atmosphere. In computer vision, imaging through a scattering medium has usually been related to dehazing~\cite{liu2019griddehazenet,jackson2020fast}, defogging~\cite{kratz2009factorizing}, underwater descattering~\cite{akkaynak2017space,sheinin2016next}, or recovering properties of a medium,  assuming its spatial uniformity~\cite{che2020towards}. 
We focus on imaging of {\em clouds}~\cite{liu2020hyperspectral,zhang2018cloudnet}. Clouds have interactions with the global climate system which are not well understood. This leads to major uncertainties in climate predictions~\cite{bony2015clouds,ceppi2017cloud,boucher2013clouds}, and thus a critical motivation to properly sense these volumetric translucent objects internally. 

Clouds are usually highly heterogeneous. Furthermore, multi-view images of clouds are governed by 3D radiative transfer (RT)~\cite{evans1998spherical,hochuli2016quantitative,pincus2009computational}: a nonlinear, recursive forward model, which expresses arbitrary multiple scattering in 3D. Inverting this model is highly complex. Common methods in remote sensing try to bypass this complexity by imposing a model~\cite{nakajima1990determination,nakajima1991determination} where clouds are horizontally uniform, infinitely broad, and RT is roughly vertical (one dimensional). This is inconsistent with nature, particularly when clouds are small. Recent work in computer vision introduced 3D {\em scattering tomography}~\cite{aides2020distributed,davis20053,gkioulekas2016evaluation,gkioulekas2013inverse,holodovsky2016situ,khungurn2015matching,levis2015airborne,levis2017multiple,levis2020multi} and proposed it as a viable path to study clouds. However, it is still slow and has not been scaled. 

Our proposed learning-based system, {\em 3DeepCT} infers 3D scattering-CT. While its results have quality which is comparable to explicit physics-based methods, it appears to run {\em five orders of magnitude faster}. Moreover, after training, it is scalable to broad cloud fields, exploiting GPU parallelism. We show how natural properties of clouds in cloud fields lead the architecture of 3DeepCT. This  includes a convolutive neural network (CNN) based on 2D convolutions, the size of its receptive field and layer-depth, and avoidance of dimensionality reduction. Furthermore, we provide an approach to train this system using rigorous physics-based generation of simulated fields and images.  

\section{Theoretical Background}
\label{sec:theorback}

We seek to recover the 3D volumetric optical parameters of a medium, particularly 3D clouds. We now provide background on inverse problems and focus on this domain.  


\subsection{Inverse Problem}
\label{sec:inv}

A 3D volumetric continuous-valued field is expressed as a vector $\boldsymbol\beta$, whose length equals the number of voxels in the domain. In our case, $\boldsymbol\beta$ represents the {\em extinction coefficient} field of a scattering medium.  There is a  forward model $\mathcal{F}(\boldsymbol{\beta})$, which renders images given $\boldsymbol{\beta}$. The measured radiance image data are expressed in a vector $\boldsymbol{y}$. Its length is the product of the number of camera image pixels, viewpoints, spectral bands and possibly polarization channels.

Estimation of $\boldsymbol\beta$ is an inverse problem. Often, it is formulated as an optimization problem:
\begin{equation}
    \boldsymbol{\beta}^* = ~\underset{\boldsymbol{\beta}}{\arg\min} ~\mathcal{E}({\boldsymbol \beta})\;,
    \label{eq:optimization_problem}
\end{equation}
where $\mathcal{E}$ is a cost function. Specifically, 
\begin{equation}
  \begin{split}
    \mathcal{E}({\boldsymbol \beta}) = ~ ||\boldsymbol{y}-\mathcal{F}(\boldsymbol{\beta})||^2_2.
    \label{eq:error}
  \end{split}
\end{equation}
In case of scattering media,  $\mathcal{F}$ is nonlinear in $\boldsymbol{\beta}$ . Hence, Eq.~(\ref{eq:optimization_problem}) cannot be solved using linear-algebra tools of matrix inversion. Nevertheless, as $\mathcal{F}$ is differentiable in $\boldsymbol\beta$, it is possible to use gradient-based methods to solve Eq.~(\ref{eq:optimization_problem}). 

Consider scattering media. Let $\boldsymbol{x}$ denote a 3D location and $\boldsymbol{\omega}$ denote a 3D direction unit vector. This vector expresses the propagation direction of radiation. The direction changes from  $\boldsymbol{\omega}'$ to $\boldsymbol{\omega}$ by scattering. This change is set by the dimensionless {\em scattering phase function} at $\boldsymbol{x}$, denoted $p(\boldsymbol{x},\boldsymbol{\omega}\cdot \boldsymbol{\omega}')$. The phase function is normalized, as it is equivalent to the probability density of scattering between directions $\boldsymbol{\omega}'$ and $\boldsymbol{\omega}$.

When radiation interacts with a particle, the radiation can be either scattered or absorbed. Scattering and absorption have, respectively, relative probabilities 
$\varpi$ and $1-\varpi$, where $0\leq \varpi\leq 1$ is the {\em single scattering albedo} of the particle. In visible light, for both air molecules and cloud water droplets, $\varpi\sim 1$, i.e., absorption is negligible. 

The extinction coefficient around 
location $\boldsymbol{x}$ is $\beta(\boldsymbol{x})$.  There, interaction of any kind 
along an infinitesimal distance ${\rm d}\boldsymbol{x}$ has 
probability $\beta(\boldsymbol{x})d\boldsymbol{x}$.
Transmittance on a straight line between two points $\boldsymbol{x}_1,\boldsymbol{x}_2$ is
\begin{align}
   T (\boldsymbol{x}_1, \boldsymbol{x}_2) =
    \exp{\left[-\int_{\boldsymbol{x}_1}^{\boldsymbol {x}_2}
	\! \beta(\boldsymbol{x}){\rm d}\boldsymbol {x}\right]}.
   \label{eq:T}
\end{align}
The richness of multiple scattering, including various paths to interaction, possible absorption and scattering events, and scattering to different angles, embodies RT. It is expressed~\cite{chandrasekhar1950radiative} by a set of recursively coupled equations:
\begin{align}
   \begin{split}
      I(\boldsymbol{x},\boldsymbol{\omega})= &I({\boldsymbol x}_0,{\boldsymbol \omega}) T\left({\boldsymbol x}_0,{\boldsymbol x}\right)  \\
    & 
    + \int_{{\boldsymbol x}_0}^{\boldsymbol x} 
       J({\boldsymbol x}',{\boldsymbol \omega})\beta({\boldsymbol x}') 
            T\left({\boldsymbol x}',{\boldsymbol x}\right) {\rm d}{\boldsymbol x}', \\
      J(\boldsymbol{x},\boldsymbol{\omega})  = 
    & \frac{\varpi({\boldsymbol x})}{4\pi}    \int_{4\pi}
      p\left({\boldsymbol x},{\boldsymbol \omega}{\cdot}{\boldsymbol \omega'}\right) 
      I\left({\boldsymbol x},{\boldsymbol \omega'}\right) 
      {\rm d}{\boldsymbol \omega}'.\\   
   \end{split}
\label{eq:RTE}
\end{align}
where $I(\boldsymbol{x},\boldsymbol{\omega})$ is the radiance field at each location and direction 
and $I({\boldsymbol x} _0,{\boldsymbol \omega})$ is input radiance to the medium in direction $\boldsymbol \omega$ at boundary point ${\boldsymbol x}_0$ (boundary condition). The field $J(\boldsymbol{x},\boldsymbol{\omega})$ is termed the {\em source function}.

Equation~(\ref{eq:RTE}) provides the radiance anywhere. Imaging by cameras samples this radiance field: then, $\boldsymbol{x}$ is the camera's center of projection, while 
$\boldsymbol{\omega}$ corresponds to a line of sight that projects to a specific pixel in the  camera at $\boldsymbol{x}$.  Hence, the {\em forward model} $\mathcal{F}$ comprises of two consecutive steps: (a) Run Eq.~(\ref{eq:RTE}), which depends on the medium $\boldsymbol{\beta}$, and (b) sample the radiance at the camera locations and lines of sight that project to pixels in these cameras. This sampling is independent of $\boldsymbol{\beta}$.

For a given medium, the RT forward model~(\ref{eq:RTE}) is computed by established methods, such as SHDOM~\cite{aides2020distributed,evans1998spherical,vshdom} and Monte-Carlo~\cite{cornet2010three,davis20053,holodovsky2016situ,mayer2009radiative}. Moreover, in recent years, approximations to the Jacobian $\partial \mathcal{F}(\boldsymbol{\beta})/\partial \boldsymbol{\beta}$ have been derived by the computer vision and graphics community \cite{loeub2020monotonicity,levis2015airborne,gkioulekas2016evaluation,gkioulekas2013inverse,luan2020langevin,9054146}. Based on the approximated Jacobian, an approximate gradient $\partial {\mathcal E}(\boldsymbol{\beta})/\partial \boldsymbol{\beta}$ is derived. This enables practical solution to the inverse problem~(\ref{eq:optimization_problem}).

However, gradient-based optimization has two main problems. First, the solution is significantly dependent on an initial guess, because $\mathcal{F}(\boldsymbol{\beta})$ is non-linear in $\boldsymbol{\beta}$ and $\mathcal{E}({\boldsymbol \beta})$ is not unimodal. Second, the problem is very difficult to scale~\cite{loeub2020monotonicity}, and it runs typically on small domains. 
Computation of $\mathcal{F}(\boldsymbol{\beta})$ and 
$\partial {\mathcal E}(\boldsymbol{\beta})/\partial \boldsymbol{\beta}$ is complex. It requires recursive evaluation of the radiance field $I$ and the source function $J$, relying on Eq.~(\ref{eq:RTE}) as well as rendering of modeled projected images, during iterated optimization. We seek to mitigate this load, by designing a learning-based system for volumetric scattering CT.

\subsection{Simulated Data and Noise} 
\label{sec:Noise}
Learning-based systems for analysis of high-dimensional data and unknowns require a large training database. It is very difficult to obtain large databases for real-world, large heterogeneous volumetric objects. For example, in clouds, which are dynamic, a real-world ground-truth training database would require, for each real-world cloud, in-situ measurements of cloud droplets, simultaneously in ${\cal O}(10^5-10^6)$ voxels, using this number of airborne cloud-droplet sensors. It is doubtful if such a large distributed ground-truth in-situ system would ever exist. Even if it will exist, it would need to be scaled to sample a large  database of clouds, in varying atmospheric and illumination conditions.  

To overcome the practical absence of real-world labeled data, we train the tomographic analysis system using meticulous simulations. We explain the simulations that relate to clouds in Sec.~\ref{sec:Data}. Here we provide background on imaging noise, which is independent of the object type, and thus applies to a variety of CT problems. We use the noise model in our simulated database. 

Section~\ref{sec:inv} explains how a forward model derives a theoretic radiance field
$I(\boldsymbol{x},\boldsymbol{\omega})$. However, real-world radiance is in the form of a random photon flux, which obeys a Poissonian distribution. The photons are converted to a discrete electric charges the sensor. Furthermore, the sensor introduces noise due to various causes, according to its specifications. Let $i^{\rm e}$ be the expected photo-electron count of a pixel. At darkness and infinitesimal exposure time, readout noise has standard deviation $\rho_{\rm read}$, in ${\rm electrons}$. A temperature $T$, the sensor dark current in $\rm{electrons/sec}$ is $D_T$. The exposure time is $\Delta t$. The standard deviation of quantization noise in ${\rm electrons}$ is $\rho_{\rm digit} = g^{\rm e}/ \sqrt{12}$, where $g^{\rm e}$ is the number of photo-electrons required to change a unit gray level~\cite{Schechner2007MultiplexingFO}. Overall, in a pixel readout, in units of ${\rm electrons}$, the noise has variance of approximately
\begin{eqnarray}
   V=i^{\rm e} + D_T\Delta t + \rho_{\rm read}^{2} + \rho_{\rm digit}^{2}
     \;.
    \label{eq:noise_var}
\end{eqnarray}

Our simulations of a perspective camera have a noise model based on the CMV4000 sensor~\cite{cmv4000}. The pixel size is $5.5 \times 5.5{\rm micron}^2$, $\rho_{\rm read} = 13$ electrons, $D_T= 125$ $\rm{electrons/sec}$ at $25^{\rm o}{\rm C}$, full well of a pixel is 13,500 ${\rm electrons}$. It uses 10bit quantization, thus $g^{\rm e} = 13,500/2^{10}$. The exposure $\Delta t$ is $762\cdot 10^{-6}~{\rm second}$. 


\subsection{Clouds in a Field} 
\label{sec:clouds}

We test learned-tomography on atmospheric clouds. This medium has characteristics that affect our system.\\

\noindent {\bf Significance of warm clouds}. Clouds account for 2/3 of Earth's albedo while warm shallow clouds constitute a major part of it, which is also the most significant part of clouds' radiative effects in the visible light for Earth's energy budget. Warm clouds are made of liquid water droplets, they are the prime scatters of sunlight, with major effects on climate. Moreover, warm clouds over the ocean have particular dramatic effect. Consider Fig.~\ref{fig:example_cloud_field}.
The ocean is dark (highly absorbing) while clouds are white (highly scattering and reflecting). Hence, the difference between cloudy and non-cloudy regions has an extreme effect on Earth's albedo and the energy  balance. Due to their significance both in the optical signal of scattering and on climate, the {\em focus of the system we demonstrate is on warm clouds over the ocean.} \\

\noindent {\bf Vertically thin, horizontally wide domain}. There are marked differences between vertical and horizontal coordinates. Atmospheric domains of interest, and cloud fields in particular, can be thousands of kilometers wide. On the other hand, the vertical dimension is thin. Atmospheric pressure (approximately the density) falls exponentially with altitude, dropping to half sea-level pressure at $\approx 5-6~{\rm km}$.
Warm clouds have tops that are typically under $\approx 2~{\rm km}$. Hence, there is a difference of {\em orders of magnitude} between horizontal and vertical lengths. Moreover, clouds are created by vertical air currents driven by gravity and buoyancy of air parcels due to temperature gradients. In contrast, horizontal wind is driven by a subtle horizontal pressure gradient, which may be null. 
For this reason, our {\em learning-based system is anisotropic, treating vertical and horizontal variations very differently}. \\

\noindent {\bf Horizontal Stationarity}.
As seen in Fig.~\ref{fig:example_cloud_field}, cloud fields maintain rather stationary statistics over long ranges. This is particularly true for clouds on the ocean, far from land. Because the statistics are approximately {\em space invariant} (in the horizontal coordinates), our analysis system is space invariant as well: this naturally leads to a {\em convolutive architecture} when treating horizontal coordinates. The vertical coordinate is separate and not space invariant.  

\section{3DeepCT}
\label{sec:3DeepCT}

Here we present a system, 3DeepCT, to learn 3D CT. Based on DNNs, 3DeepCT learns and then infers volumetric media. {\em 3DeepCT is faster by orders of magnitude}, relative to pure physics-based scene analysis. Its 
input is multi-view images of the scene, and its output is an estimated 3D heterogeneous  $\hat{\boldsymbol{\beta}}$. As described in Sec.~\ref{sec:Noise}, it is extremely challenging, to say the least, to empirically acquire a large training database for large heterogeneous volumetric media, which are typically dynamic. Hence, training is based on rigorous physics-based models of the true scenes 
 $\{ \boldsymbol{ \beta}^{\text{true}} \}$, images and realistic noise. Using such a simulated database, learning is performed by minimizing a loss. The loss we use is the mean square error (MSE):
\begin{align}
    {\rm Loss}_{{\rm MSE}}(\boldsymbol{ \beta}^{\text{true}},\hat{\boldsymbol{\beta}})=
    ||\boldsymbol{\beta}^{\text{true}}-\hat{\boldsymbol{\beta}}||^2_2
    \label{eq:mse_loss}
\end{align}

3DeepCT is based on physics-based training data, hence in the analysis of test data, physics plays an implicit role. However, for scientific needs, it benefits that the final word is given to an explicit physics-based solution to the inverse-problem. A {\bf hybrid system} can do this. Given a new data of an unknown scene, 3DeepCT can provide a solution which is comparable to explicit iterated physics-based inversion, yet doing so very fast. The solution obtained by  3DeepCT is then used as initialization for the much slower physics-based inversion. Only a few iterations would then be needed for explicit physics-based inversion. 

These principles can apply to any type of computed tomography and any media. However, it benefits an expert system to be tailored to an imaging modality and the media being observed. Therefore, from this point on, we describe how the architecture and implementation of 3DeepCT is tailored to scattering CT of warm clouds. 

\subsection*{Architecture Tailored to Cloud Fields}
\label{sec:Architecture}

These architectural principles follow the natural traits of clouds, some of which are described in Sec.~\ref{sec:clouds}:\\

\noindent $\{{\rm {\tt 1}}\}$ Due to the horizontal stationarity of cloud fields, both as volumetric objects and in images, 3DeepCT is a convolutive neural network (CNN). Convolution kernels are operated in the horizontal domain.\\

\noindent  $\{{\rm {\tt 2}}\}$  Data is in the form of multi-angular images, ie., the input has {\em angular channels} corresponding to different viewpoints (view angles). To capture 3D information from multiview images, all angular channels participate. There is no preference to angular-neighborhood, contrary a convolutive operation. Moreover, output clouds have a vertical structure expressed by {\em vertical channels}. As written in Sec.~\ref{sec:clouds}, cloud vertical structure is not stationary. Therefore in all layers of the DNN, convolution kernels operate {\em only} in 2D. The channels on the complementary coordinate, specifically angular channels in the input and vertical channels in the output, are {\em completely intertwined, per neural network layer}. \\

\noindent  $\{{\rm {\tt 3}}\}$  Due to the effective thinness of the atmospheric domain (Sec.~\ref{sec:clouds}) and typically small number of viewpoints, having a completely intertwined architecture 
in the vertical or angular direction is not a computational burden. On the other hand, due to the very wide extent of cloud fields and high resolution of imagers, a horizontal CNN makes analysis scalable. Principles  $\{{\rm {\tt 1,2}}\}$ are addressed directly by using the {\em Conv2D} operation at entry stage of any layer of the 3DeepCT neural network. \\

\noindent  $\{{\rm {\tt 4}}\}$ A small warm cloud cell typically has size of hundreds of meters to a kilometer. Thus an output neuron needs to have some statistical dependency to the output of other neurons in that range. This dictates the minimum {\em receptive field} range $R$ of the overall DNN. This implies the DNN {\em depth}, i.e, its number of layers. 
Let the camera horizontal resolution be $r$ meters/pixel. In principle $\{{\rm {\tt 3}}\}$, operations per layer are based on {\em Conv2D}, which uses kernels having a small 2D support, of length $h$ pixels. The number of layers needs to be $L\sim R/(rh)$.\\   

\noindent  $\{{\rm {\tt 5}}\}$ Usually, clouds are created by chaotic turbulent flow. Thus they are  
not highly structured objects as the human body. Consequently, clouds are not amenable to significant dimensionality reduction. Moreover, tomography tries to resolve voxels, whose number is comparable to the data size. Therefore, contrary to auto-encoder networks, we do not include dimensionality reduction (pooling) operations.

\begin{figure*}[t]
\begin{center}
   \includegraphics[trim=0cm 0cm 0cm 0cm,clip, width=0.9\linewidth]{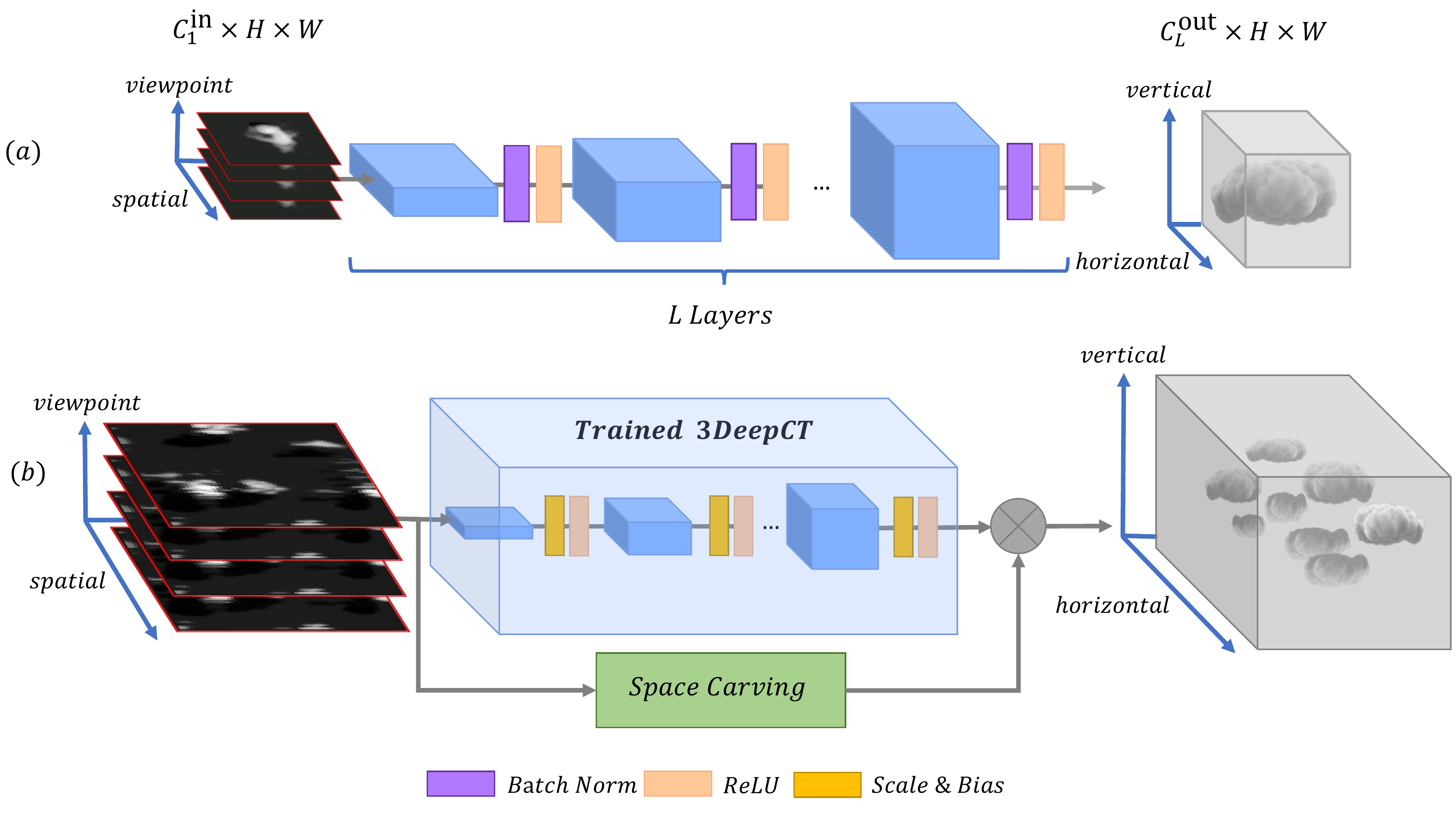}
\end{center}
   \caption{{\bf 3DeepCT neural network architecture:} A convolutional neural network receives multi-view satellite-images of a cloud as input and reconstructs the 3D extinction field of the cloud. (a) The architecture we use to train the network. The images are of single isolated clouds. (b) Illustration of the architecture we use for inference, where we use large images that captures a larger cloud field. A space-carving mask is incorporated. Batch normalization layer degenerates to linear operations of scaling and bias.}
\label{fig:net_architecture}
\end{figure*}

The architecture is illustrated in Fig.~\ref{fig:net_architecture}.
3DeepCT has $L$ layers, indexed $l=1\ldots L$. In each layer, there are, respectively $C^{\rm in}_l$ and $C^{\rm out}_l$ input and output channels, where $C^{\rm in}_{l}=C^{\rm out}_{l-1}$ for $l\neq 1$.
The number of viewpoints sets $C^{\rm in}_{1}$, while
$C^{\rm out}_{L}$ is the number of resolved vertical altitudes in the volume. Each input image (data per input channel) is of size $H\times W$ pixels. Since we have no dimensionality reduction, each layer in 3DeepCT maintains a 2D size of $H\times W$, and this is the horizontal size of the estimated volume domain. Moreover, 
 $C^{\rm out}_l\geq C^{\rm in}_l$, gradually increasing (if not maintaining) the number of channels per layer.  

During testing, i.e, analysis of an unknown scene, each layer $l$ is composed of two operations, Conv2D, followed by an element-wise ReLU function. Let the content of channel 
$c=1\ldots C^{\rm in}_{l}$ be the  $H\times W$ array 
 ${\bf A}_{(l,c)}$. It undergoes convolution with a corresponding 2D kernel denoted ${\bf W}_{(l,c,c')}$, 
where $c'=1\ldots C^{\rm out}_{l}$. That is, the same layer
is independently filtered by $C^{\rm out}_{l}$ kernels. Then, the output of layer $l$, i.e the input of layer $l+1$, per channel $c'$ is
 \begin{equation}
    {\bf A}_{(l+1,c')} 
    = {\rm ReLU}
    \left\{
        {\bf B}_{(l,c')} 
        +
        \sum_{c=1}^{C^{\rm in}_{l}} 
        {\bf A}_{(l,c)} 
        \star
        {\bf W}_{(l,c,c')}
    \right\}.
    \label{eq:convlayer}
\end{equation}
Here $\star$ denotes 2D convolution and ${\bf B}_{(l,c')}$
is a  $H\times W$ array of scalars (biases), per layer and channel $l,c'$.

Training of the DNN optimizes 
${\bf W}_{(l,c,c')},{\bf B}_{(l,c')}$, $\forall l,c,c'$.
During training, known ground-truth examples are introduced in batches of $b$ scenes. While training on a batch, {\em batch-normalization} \cite{ioffe2015batch} is operated prior to the ${\rm ReLU}$ operation. Recall from Sec.~\ref{sec:Noise} the practical infeasibility of obtaining an empirical real-world database of clouds. Thus, 3DeepCT  trains using rigorous physical simulations, described in Sec.~\ref{sec:Data}. 

We found that testing significantly improved by {\em space-carving}~\cite{veikherman2014clouds} of the volumetric domain. Clouds have significant contrast relative to the dark ocean background. Hence, a moderate cloud-mask is applied per input image, followed by space carving. This yields a simple bounding constraint on which voxels may potentially have water droplets.
 
We implemented 3DeepCT using the Pytorch framework~\cite{paszke2019pytorch}. We use $L=33$. Training uses $b=32$, 1000 epochs and $H=W=32$. Due to the convolution architecture, testing is enabled at arbitrarily larger $H,W$. Schemes of the network train and test architectures shown in Fig.~\ref{fig:net_architecture}.

\section{Data} 
\label{sec:Data}

\subsection{Physics-based Clouds} 
\label{sec:Clouds_Formation}

We train 3DeepCT to recover the 3D extinction coefficient field of warm clouds in the atmosphere. This is done using careful simulations. The simulations couple two components. One is the core (dynamical) large eddy simulation (LES) model, which solves the coupled equations of a turbulent atmosphere. LES is the main numerical tool for generating and studying clouds at the altitudes of relevance to our work~\cite{heus2009statistical,neggers2003size,xue2006large}. Here we use the System for Atmospheric Modeling (SAM) ~\cite{khairoutdinov2003cloud}. SAM is a non-hydrostatic, inelastic model. 

The second component handles the droplet's microphysics, by a spectral (bin) model (HUJI SBM ~\cite{khain2004simulation,fan2009ice}). It explicitly evolves physical equations of the processes that affect cloud droplet growth, thus yielding the size distribution of droplets per voxel. This distribution is sampled into 33 bins, logarithmically spread in the range $[2~{\rm \mu m},3.2~{\rm mm}]$. For this distribution we calculate the effective radius and effective variance of the cloud's droplets in each voxel, and for computational efficiency we limit them to the maximal values of $20~{\rm \mu m}$ and 2 respectively.
This information yields the ground-truth optical extinction coefficient $\beta^{\text{true}}$ per voxel, through Mie theory~\cite{grainger2004calculation}. 

The cloud field was simulated using the BOMEX~\cite{siebesma2003large} setup which is based on surface fluxes, large-scale forcing, and profiles of wind, humidity, and temperature that had been measured in a prior field campaign in Barbados. 
The simulated domain is $12.82~{\rm km}\times 12.82~{\rm km}$ wide, with cyclic horizontal boundary conditions. The simulated clouds have spatial resolution of $50~{\rm m}$ horizontally and $40~{\rm m}$ vertically. Time duration of the simulation data is 8 hour (including 2 hours of  spin-up time), and we use a snapshot every 2 minutes to produce the database of cloud fields. The simulation evolved in 1 second increments, each yielding a different 3D spatial field which includes dozens of clouds. 

The data created  takes 1.2 TB of memory.  It took approximately 5 days to generate, on an Intel Xeon Gold 5115 with 256 cores. From this data, we created a main database of cropped 4800 training samples, 1200 evaluation samples and 286 test samples. We also created an auxiliary database, in which the liquid water content of the same clouds is reduced by a factor of 25, to enable representation and network expression of smaller optical depths. The auxiliary database consists of $3788$ training samples, $947$ evaluation samples and $500$ test samples. All sample volumes contain isolated clouds in domains having  $1.6 {\rm km}\times  1.6{\rm km}$ horizontal extent and $1.28~{\rm km}$ vertical extent.

\subsection{Physics-based Rendering} 
\label{sec:Images_Formation}

Given the ground-truth clouds, we render images using a physics-based RT equation solver (SHDOM), which has an open-source code~\cite{pyshdom_code}. Rendering is set per solar angle, multi-view geometry and sensor characteristics. 
We trained 3DeepCT using a NVIDIA GeForce RTX 2080 computer. We separately trained versions of 3DeepCT for several distinct imaging geometries:

\noindent {\bf 32 Viewpoints}. A northbound string-of-pearls \cite{kleinschrodt2016comparison} formation of 32 satellites orbit at $600{\rm km}$ altitude. Nearest-neighbor satellites are $100{\rm km}$ apart. 
They view the same field in off-nadir angles 
$-75.2^\circ$,
$\pm73.5^\circ$,
$\pm71.7^\circ$,
$\pm69.6^\circ$,
$\pm67.4^\circ$,
$\pm64.8^\circ$,
$\pm62^\circ$,
$\pm58.8^\circ$,
$\pm55.2^\circ$,
$\pm51^\circ$,
$\pm46^\circ$,
$\pm40.6^\circ$,
$\pm34^\circ$,
$\pm26^\circ$,
$\pm18^\circ$,
$\pm9^\circ$,
and $0^\circ$. 
Each carries a perspective camera. The field of view of each camera is $0.22^\circ$, corresponding to a ground footprint at the nadir of $1.6{\rm km}\times1.6{\rm km}$, at $50~{\rm m}$ ground resolution. The sensor noise characteristics correspond to the CMV4000 sensor \cite{cmv4000}. Solar azimuth and zenith angles are $45^{\circ}$ and $30^{\circ}$, respectively. Here, the 3DeepCT architecture  has $C^{\rm in}_{1}=C^{\rm out}_{L}=32$.  We trained this model on the main database for 1000 epochs, which took $\approx 11~{\rm hours}$.

\noindent {\bf 10 Viewpoints}. This geometry is motivated by the CloudCT space mission~\cite{schilling2019cloudct}. This mission plans 10 nanosatellites carrying perspective cameras, which will simultaneously image clouds in a multiview geometry. This geometry is visualized in Appendix \ref{sec:app_10_viewpoints}. 
The parameters are similar to those of the {\em 32 Viewpoints} geometry. Here, however, we use off-nadir angles $-46^\circ$,
$\pm34^\circ$,
$\pm26^\circ$,
$\pm18^\circ$,
$\pm9^\circ$ and
$0^\circ$, 
and the architecture has $C^{\rm in}_{1}=10, C^{\rm out}_{L}=32$. 
We trained this model on the main database for 1000 epochs, which took $\approx 8~{\rm hours}$.


\section{Simulated Inference Results} 
\label{simulResults}

We present inference (testing) results and compare four approaches:\\
\noindent {\tt A.}  { \bf 3DeepCT}. It runs on a GeForce RTX 2080, where it takes just {\em milliseconds} to reconstruct each cloud. This enables us to test hundreds of cloud samples. \\ 
\noindent {\tt B.}  { \bf Physics-based inverse scattering}. This explicit optimized iterated fit of a physical model to the data is the state of the art~\cite{levis2015airborne,levis2017multiple,levis2020multi}. It uses an SHDOM code in the public domain~\cite{pyshdom_code}. It is initialized by a default constant value of ${\boldsymbol \beta}^{\rm initial}=0.01$ globally for all the cloud's space-carving mask. To run this method we use 20 cores of Intel(R) Xeon(R) Gold 6240 CPU @ 2.60GHz with 72 cores. The state-of-the-art approach
does not exploit a GPU, as its complex calculations require CPUs. It takes about an  {\em hour} to reconstruct each cloud in this system. Hence, when performing cross-method comparisons, we used a {\em subset of seven} specific clouds from the test dataset (shown in Appendix \ref{sec:app_simulated_inference_results}). \\
\noindent {\tt C.} { \bf Hybrid system}. The result of {\tt A} provides an initialization to subsequent use of approach {\tt B}. Then, iterations explicitly optimize physics-based inverse scattering~\cite{pyshdom_code} iteratively, as in {\tt B}.\\
\noindent {\tt D.} { \bf Quick hybrid system} This is similar to {\tt C}, but only 10 iterations are performed, leading to a shorter overall run time. 

\begin{figure*}[t]
    \begin{center}
      \includegraphics[trim=2cm 10cm 0cm 4cm,clip, width=1\linewidth]{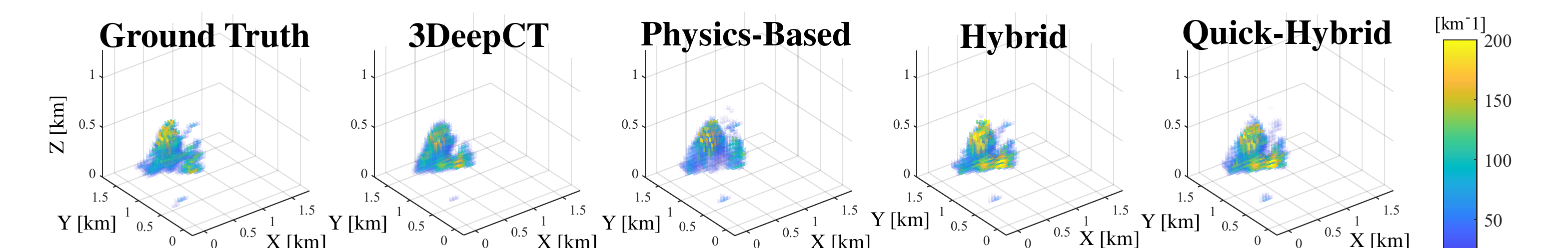}
    \end{center}
   \caption{3D reconstructions of cloud extinction. These recovery results correspond to an example cloud out of the {\em subset of seven} clouds tested. From left to right: 3D ground truth extinction of the cloud; 3D reconstructed extinction using four methods mentioned in Sec. \ref{simulResults}: 3DeepCT, Physics-Based Inverse Scattering, Hybrid system, Quick Hybrid system.}
    \label{fig:4_methods_beta}
\end{figure*}
\begin{figure}
\begin{center}
   \includegraphics[trim=0.3cm 9cm 19.5cm 0.3cm,clip, width=1\linewidth]{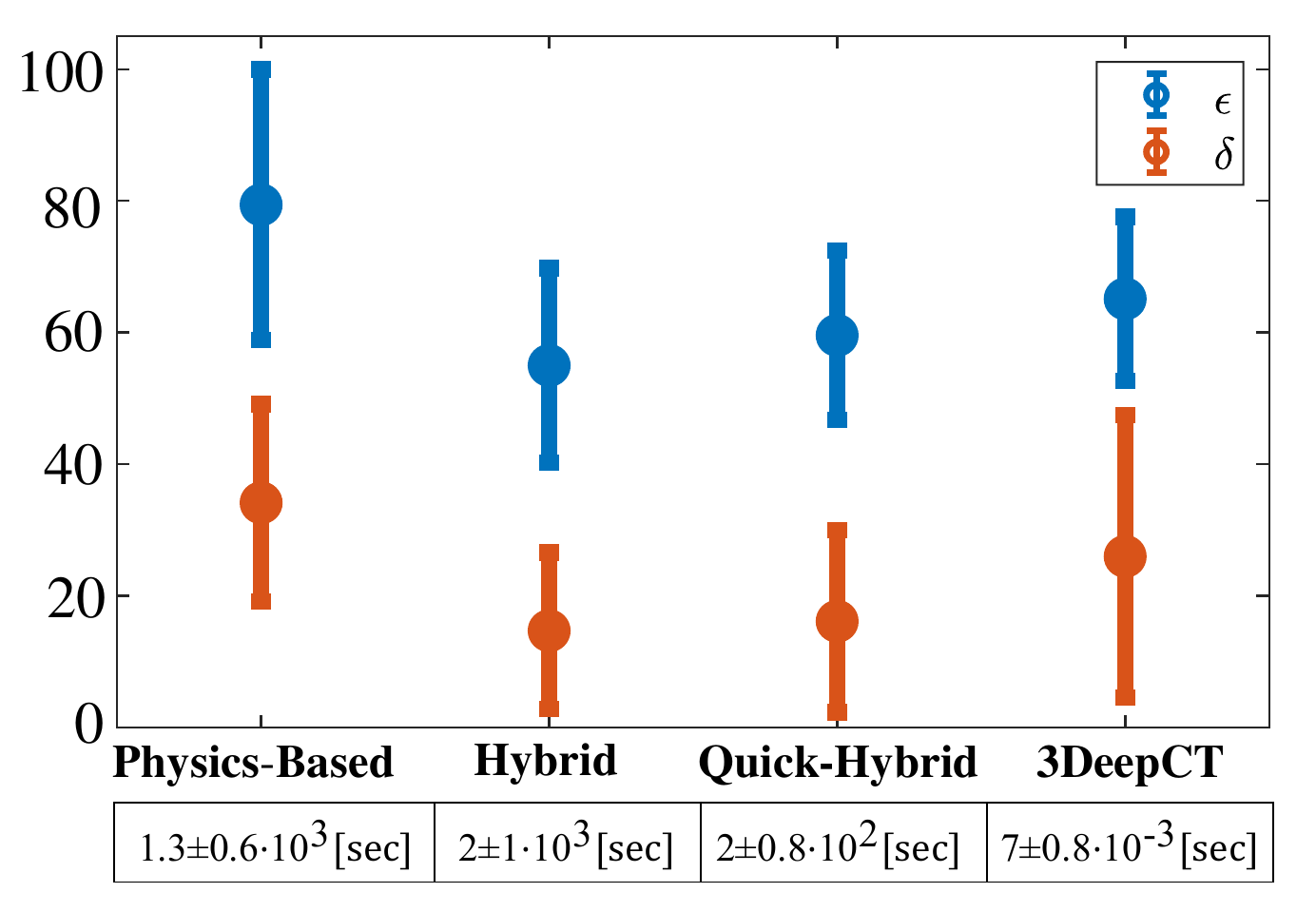}
\end{center}
   \caption{Numerical results of the {\em subset of seven} clouds, discussed in Sec. \ref{simulResults}. Blue is $\epsilon$ and orange is $\delta$ (see Eq.~(\ref{eq:epsilon_delta})). The circles are the mean values and the lines express $\pm$ standard deviation (std) of the results.}
\label{fig:7_clouds_box_plot}
\end{figure}

As example, recovery of one of the clouds in the said {\em subset of seven} using the {\bf 32 Viewpoints} geometry is shown in Figs.~\ref{fig:4_methods_beta} and~\ref{fig:4_methods_scatter_plot} (additional results shown in Appendix \ref{sec:app_simulated_inference_results}).
\begin{figure}[t]
    \begin{center}
      \includegraphics[trim=3cm 7.5cm 5cm 7.3cm,clip,    width=1\linewidth]{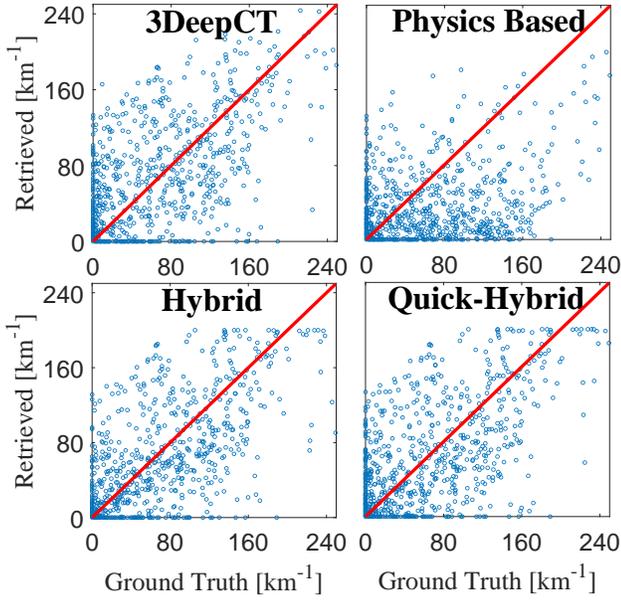}
    \end{center}
    \caption{Scatter plots of recovery results of an example cloud out of the {\em subset of seven} clouds tested. The plot related to the four methods described in Sec. \ref{simulResults}. 
    The reconstructed 3D extinction is $\hat{\boldsymbol{\beta}}$.
    The red line represents ideal reconstruction, where $\hat{\boldsymbol{\beta}}=\boldsymbol{\beta}_{\text{true}}$. Voxels deep in the cloud are difficult to reconstruct, therefore we see the scatter around the red line.}
    \label{fig:4_methods_scatter_plot}
\end{figure}
We report numerical comparisons to the ground truth $\boldsymbol\beta^{\text{true}}$. To be able to compare to prior art~\cite{levis2015airborne,levis2017multiple,levis2020multi}, we use these respective criteria for relative average error and relative total mass error. 
\begin{equation}
    \epsilon=\frac{||\boldsymbol\beta^{\text{true}}-\hat{\boldsymbol{\beta}}||_1}{||\boldsymbol{\beta}^{\text{true}}||_1},~~
    \delta=\frac{||\boldsymbol{\beta}^{\text{ true}}||_1-||\hat{\boldsymbol{\beta}}||_1}{||\boldsymbol{\beta}^{\text{true}}||_1}
    \;.
    \label{eq:epsilon_delta}
\end{equation}
The run-time and quality results relating to the {\em subset of seven} clouds using the {\bf 32 Viewpoints} geometry are summarized in Fig.~\ref{fig:7_clouds_box_plot}.
Clearly, in these test cases, 3DeepCT has a huge advantage in terms of runtime: it is about {\em five orders of magnitude faster} than the state of the art, which is a physics-based explicit approach. In terms of quality, 3DeepCT also yields, on average, better results than the state of the art. The Hybrid method yields a significant improvement of quality, relative to the state of the art, for the same run-time. The Quick-Hybrid method is a compromise or run-time and quality between 3DeepCT and a Hybrid long-run.  
 
When not comparing to slow physics-based optimization, 3DeepCT can be assessed on the full test dataset. Results are summarised in Table~\ref{table:test_results_sats_model_average_table}. There is a degradation in quality when the number of viewpoints decreases. However, a 10-satellite geometry is more realistic in the short term, as planned by the CloudCT space mission \cite{schilling2019cloudct}.
\begin{table}
    \begin{center}
    \begin{tabular}{|c|c|c|c|}
    \hline
    Model & $\epsilon$ & $\delta$ & Time [millisec] \\
    \hline\hline
    32 satellites & 82$\pm$10\% & 32$\pm$16\% & $7 \pm 0.9$\\
    10 satellites & 86$\pm$10\% & 44$\pm$16\%  & $7 \pm 0.7$\\
    \hline
    \end{tabular}
    \end{center}
    \caption{Summary of 3DeepCT test results: Mean $\pm$ standard deviation (std) of different models described in Sec. \ref{sec:Images_Formation}. Equations for $\epsilon$ and $\delta$ are in Eq. (\ref{eq:epsilon_delta}).}
\label{table:test_results_sats_model_average_table}
\end{table}

\begin{figure}[t]
\begin{center}
   \includegraphics[trim=0cm 24cm 39.5cm 0.5cm,clip, width=1\linewidth]{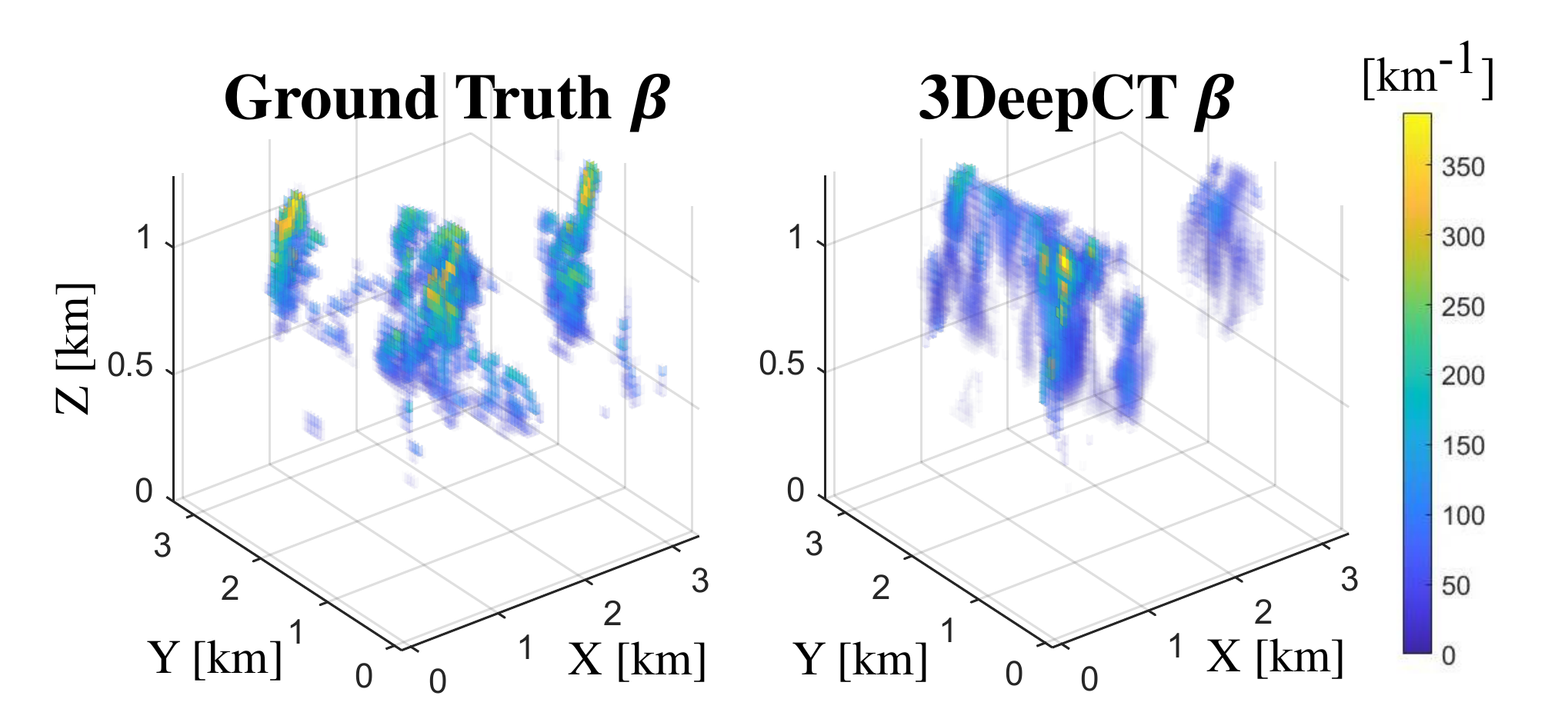}
\end{center}
   \caption{3D extinction in a large cloud field, $3{\rm km}$ long.  [Left] The true medium. [Right] Medium recovered by 3DeepCT.}
\label{fig:bigger_clouds}
\end{figure}

We also show that 3DeepCT system can be used on a larger field than it had trained on, thanks to its convolutive architecture. The result shown in Fig.~\ref{fig:bigger_clouds} is on a field whose area is four times the area used during training. Recovery here uses the {\bf 32 Viewpoints} geometry. Here $\epsilon=1.4,\delta=0.3$ and runtime took 1.7 seconds. While the CNN scales, we could not run the physics-based explicit optimization (state of the art) on this field, because it requires excessive resources, problem also reported in \cite{loeub2020monotonicity}. 

\section{Summary}
\label{sec:summary}
In this paper we presented 3DeepCT, a DNN for computed tomography, which performs 3D reconstruction of big translucent objects, where we focused on cloud fields. We showed that there is a significant advantage for this approach. In particular, it achieves improved accuracy and dramatic reduction in run time compared to physics-based methods. Thus, 3DeepCT enables the reconstruction of large cloud fields. In addition, this paves the ways for many future research directions. For instance, enriching the dataset with more diverse data will allow the system to continuously improve. Moreover, this approach encourages the development of transfer-learning techniques This will allow previously trained models to be reconfigured to new sun angles, viewpoint directions and imaging systems. We believe that there are additional scientific domains \cite{gumbel2020mats} which can benefit from our approach, allowing to successfully solve complex tasks where physics-based datasets exist and the computation complexity of state-of-the-art reconstruction is high.

\section*{Acknowledgements}
We are grateful to Ilan Koren, Orit Altaratz and Roi Ronen for useful discussions and good advice.
We thank Aviad Levis and Jesse Loveridge for the pySHDOM code and for being responsive to questions about it. We thank Johanan Erez, Ina Talmon and Daniel Yagodin for technical support. Yoav Schechner is the Mark and Diane Seiden Chair in Science at the Technion. He is a Landau Fellow - supported by the Taub Foundation.
His work was conducted in the Ollendorff Minerva Center. Minvera is funded through the BMBF.
This project has received funding from the European Union’s Horizon 2020 research and innovation programme under grant agreement No 810370-ERC-CloudCT.

\appendix
\section*{Appendix}

\section{10 Viewpoints Geometry}
\label{sec:app_10_viewpoints}
\begin{figure}[t]
   \begin{center}
   \includegraphics[trim=0cm 0cm 0cm 0cm,clip, width=1\linewidth]{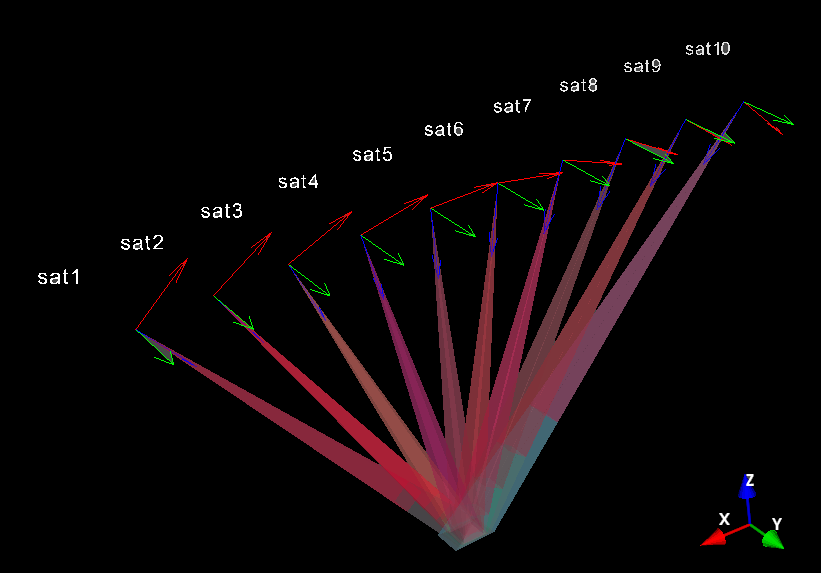}
   \end{center}
   \caption{{\bf 10 Viewpoints} geometry; orbit at $600{\rm km}$ altitude, nearest-neighbor satellites are $100{\rm km}$ apart, remotely sensing clouds in the atmosphere from space.}
\label{fig:satellites_setup}
\end{figure}

The {\bf 10 Viewpoints} geometry, presented in Sec.~\ref{simulResults} is visualized in Fig.~\ref{fig:satellites_setup}. 
Recall that this geometry uses 10 satellites orbit at $600{\rm km}$ altitude. Nearest-neighbor satellites are $100{\rm km}$ apart. 
They view the same field in off-nadir angles 
$-46^\circ$,
$\pm34^\circ$,
$\pm26^\circ$,
$\pm18^\circ$,
$\pm9^\circ$ and
$0^\circ$.
Each carries a perspective camera. The field of view of each camera is $0.22^\circ$, corresponding to a ground footprint at the nadir of $1.6{\rm km}\times1.6{\rm km}$, at $50~{\rm m}$ ground resolution. Solar azimuth and zenith angles are $45^{\circ}$ and $30^{\circ}$, respectively. Here, the 3DeepCT architecture  has $C^{\rm in}_{1}=10, C^{\rm out}_{L}=32$.  We trained this model on the main database for 1000 epochs, which took $\approx 8~{\rm hours}$. 
\begin{figure}[t]
    \begin{center}
      \includegraphics[trim=0cm 6.7cm 22cm 0.5cm,clip, width=1\linewidth]{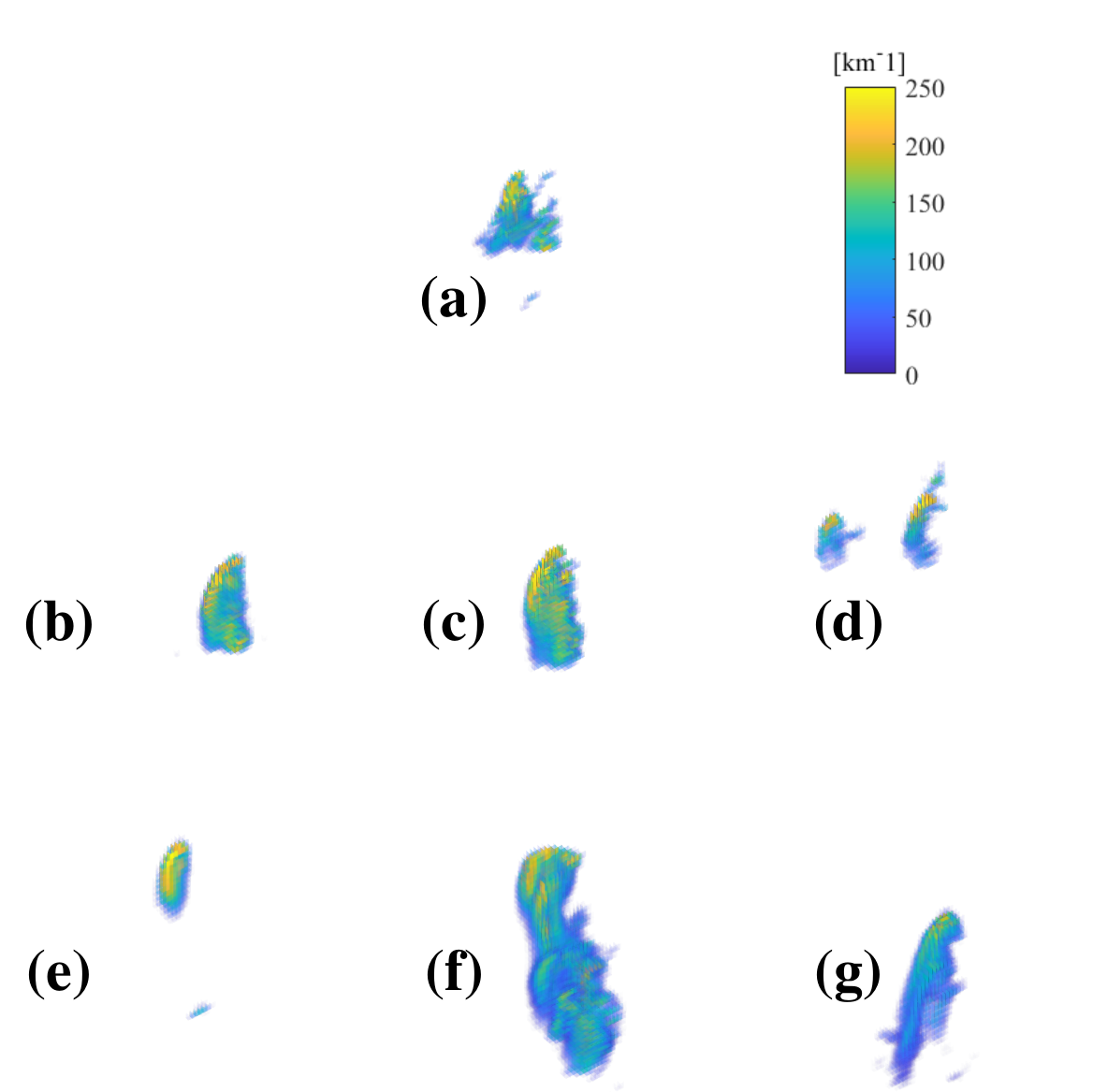}
    \end{center}
   \caption{The {\em Subset of seven} clouds.}
    \label{fig:clouds_collection_alternative}
\end{figure}

\section{Additional Simulated Inference Results}
\label{sec:app_simulated_inference_results}
\begin{figure*}[t]
    \begin{center}
      \includegraphics[trim=0cm 7cm 4cm 0cm,clip, width=1\linewidth]{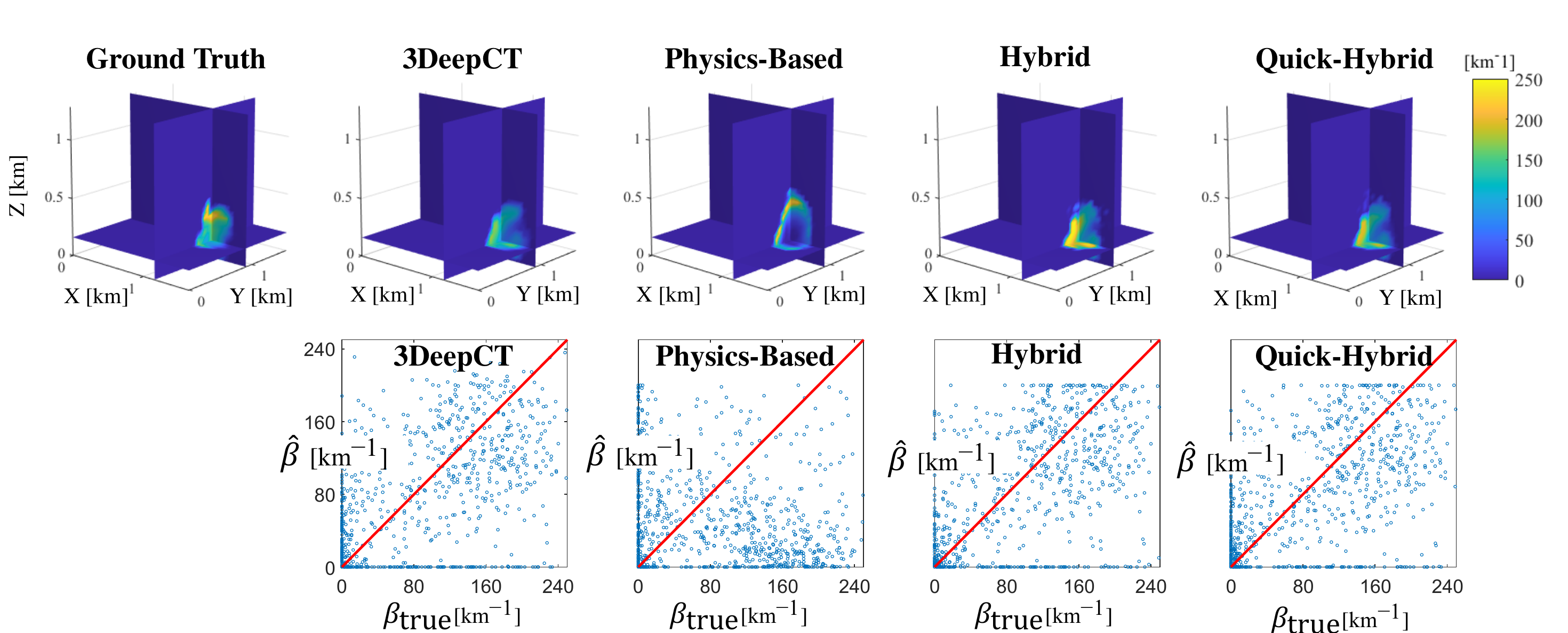}
    \end{center}
   \caption{3D reconstructions of cloud extinction. These recovery results correspond to cloud (b) out of the {\em subset of seven} clouds tested shown in Fig.~\ref{fig:clouds_collection_alternative}. [First row] From left to right: 3D ground-truth extinction of the cloud; 3D reconstructed extinction using the four methods mentioned in Sec. \ref{simulResults}: 3DeepCT, Physics-Based Inverse Scattering, Hybrid system, Quick-Hybrid system. [Second row] Scatter plots of the recovery results. The plots relate to the four methods at the top row. 
    The reconstructed 3D extinction is $\hat{\boldsymbol{\beta}}$.
    The red line represents ideal reconstruction, where $\hat{\boldsymbol{\beta}}=\boldsymbol{\beta}_{\text{true}}$.}
    \label{fig:cloud_10}
\end{figure*}

\begin{figure*}[t]
    \begin{center}
      \includegraphics[trim=0cm 7cm 4cm 0cm,clip, width=1\linewidth]{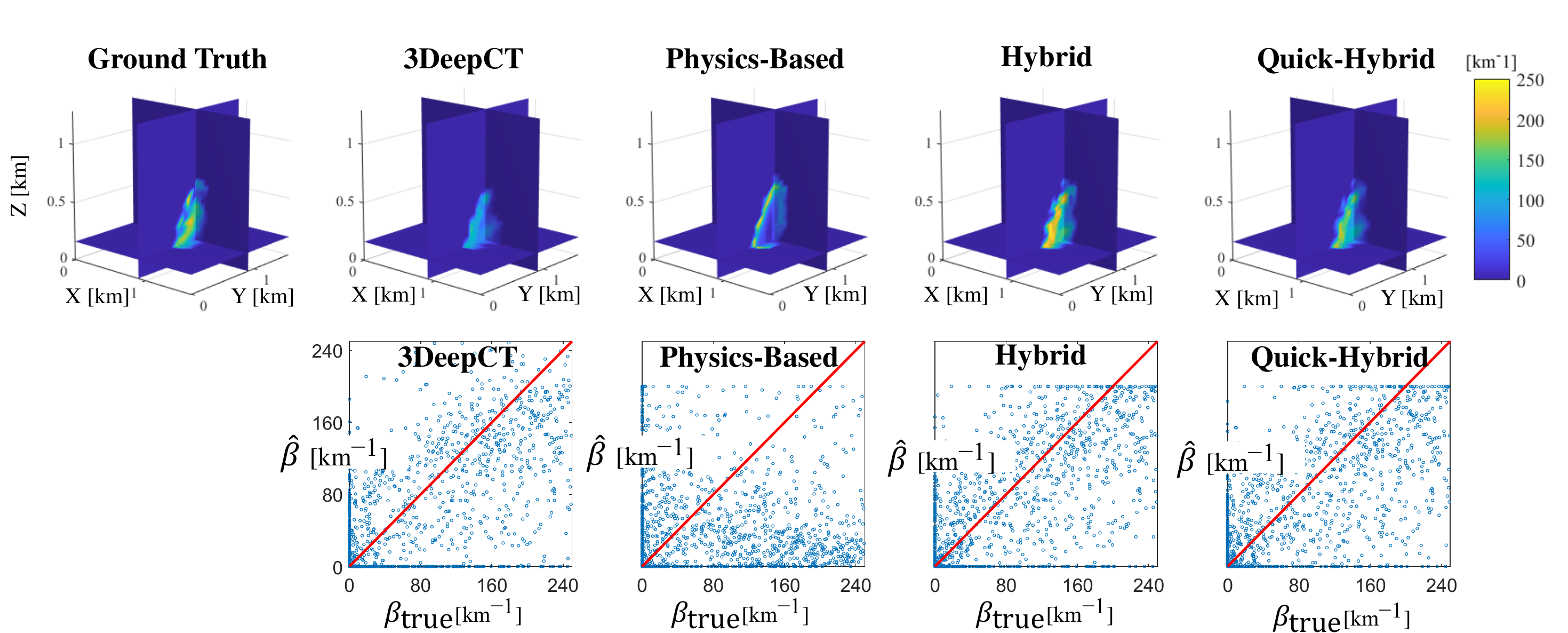}
    \end{center}
   \caption{3D reconstructions of cloud extinction. These recovery results correspond to cloud (c) out of the {\em subset of seven} clouds tested shown in Fig.~\ref{fig:clouds_collection_alternative}. [First row] From left to right: 3D ground-truth extinction of the cloud; 3D reconstructed extinction using the four methods mentioned in Sec. \ref{simulResults}: 3DeepCT, Physics-Based Inverse Scattering, Hybrid system, Quick-Hybrid system. [Second row] Scatter plots of the recovery results. The plots relate to the four methods at the top row. 
    The reconstructed 3D extinction is $\hat{\boldsymbol{\beta}}$.
    The red line represents ideal reconstruction, where $\hat{\boldsymbol{\beta}}=\boldsymbol{\beta}_{\text{true}}$.}
    \label{fig:cloud_37}
\end{figure*}

\begin{figure*}[!t]
    \begin{center}
      \includegraphics[trim=0cm 7cm 4cm 0cm,clip, width=1\linewidth]{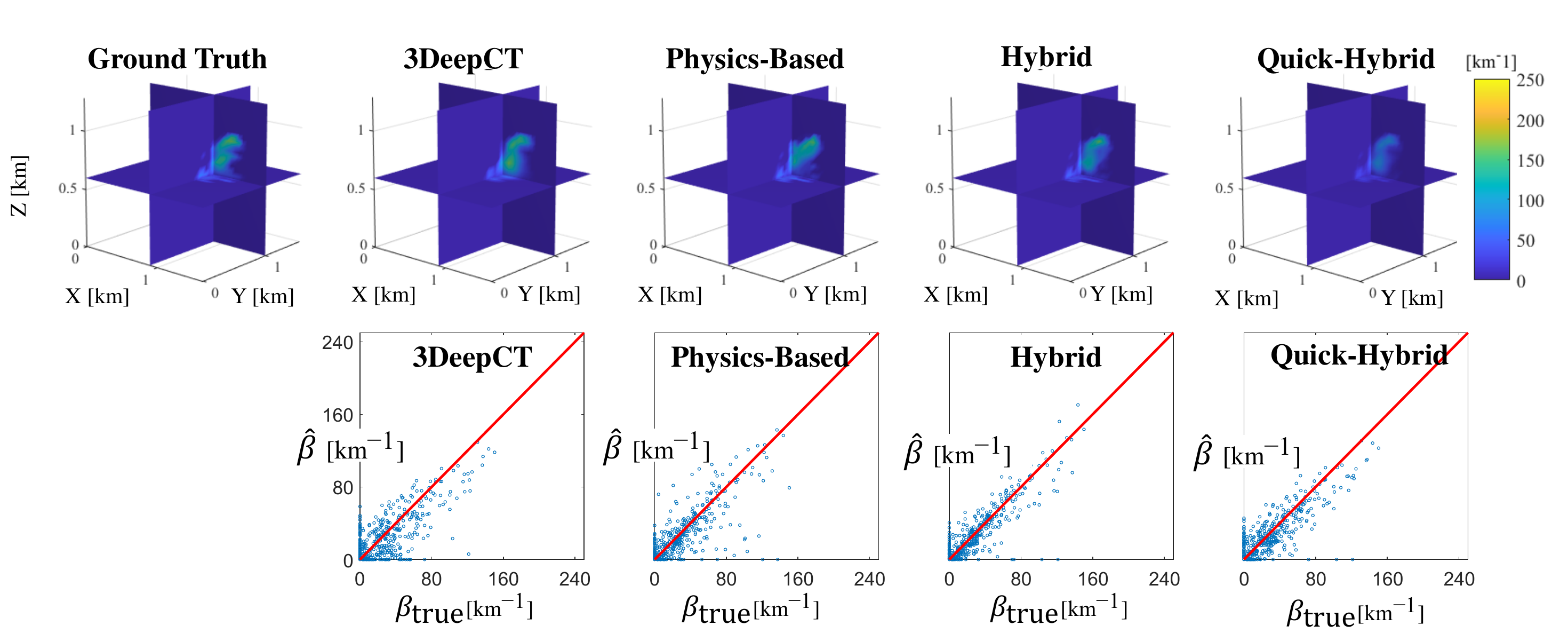}
    \end{center}
   \caption{3D reconstructions of cloud extinction. These recovery results correspond to cloud (d) out of the {\em subset of seven} clouds tested shown in Fig.~\ref{fig:clouds_collection_alternative}. [First row] From left to right: 3D ground-truth extinction of the cloud; 3D reconstructed extinction using the four methods mentioned in Sec. \ref{simulResults}: 3DeepCT, Physics-Based Inverse Scattering, Hybrid system, Quick-Hybrid system. [Second row] Scatter plots of the recovery results. The plots relate to the four methods at the top row. 
    The reconstructed 3D extinction is $\hat{\boldsymbol{\beta}}$.
    The red line represents ideal reconstruction, where $\hat{\boldsymbol{\beta}}=\boldsymbol{\beta}_{\text{true}}$.}
\label{fig:cloud_66}
\end{figure*}

Fig.~\ref{fig:clouds_collection_alternative} shows the {\em subset of seven} clouds. 
Recall that our {\bf 32 Viewpoints} model is demonstrated on one cloud out of the {\em subset of seven} clouds discussed in Sec. \ref{simulResults}. This cloud is shown in Fig.~\ref{fig:clouds_collection_alternative}(a).
The reconstruction of three additional clouds (Fig.~\ref{fig:clouds_collection_alternative}(b)-(d)) using the four methods described in Sec. \ref{simulResults} are shown in Figs.~\ref{fig:cloud_10},\ref{fig:cloud_37},\ref{fig:cloud_66} respectively.

{\small
\bibliographystyle{ieee_fullname}
\bibliography{egbib}
}

\end{document}